\documentstyle[twoside,fleqn,espcrc2,epsfig]{article}
\title{U(1) staggered Dirac operator and random matrix
  theory\thanks{Partially funded by the Department of Energy under
    contracts DE-FG02-97ER41022 and DE-FG05-85ER2500.}}
\author{Bernd A. Berg\address{Department of Physics, The Florida State
    University, Tallahassee, FL~32306, USA and Supercomputer
    Computations Research Institute, The Florida State University,
    Tallahassee, FL~32306, USA},
  Harald Markum\address{Institut f\"ur Kernphysik, Technische
  Universit\"at Wien, A-1040 Vienna, Austria},
  Rainer Pullirsch$^{\rm b}$,
  and
  Tilo Wettig\address{Institut f\"ur Theoretische Physik,
  Technische Universit\"at M\"unchen, D-85747 Garching, Germany}}
\begin{document}
\begin{abstract}
  We investigate the spectrum of the staggered Dirac operator in 4d
  quenched U(1) lattice gauge theory and its relationship to random
  matrix theory.  In the confined as well as in the Coulomb phase the
  nearest-neighbor spacing distribution of the unfolded eigenvalues is
  well described by the chiral unitary ensemble. The same is true for
  the distribution of the smallest eigenvalue and the microscopic
  spectral density in the confined phase. The physical origin of the
  chiral condensate in this phase deserves further study.
\end{abstract}
\date{\today}
\maketitle

By now it is a well-known fact that the spectrum of the QCD Dirac
operator
\begin{equation} \label{Dirac_Op}
i D + im = \left( \matrix{ im & T \cr T^{\dagger} & im \cr}
\right)\ ~{\rm in\ a\ chiral\ basis}
\end{equation}
is related to universality classes of random matrix theory (RMT),
i.e. determined by the global symmetries of the QCD partition
function~\cite{ShVe92}.  In RMT the matrix $T$ in
Eq.~(\ref{Dirac_Op}) is replaced by a random matrix with appropriate
symmetries, generating the chiral orthogonal (chOE), unitary (chUE),
and symplectic (chSE) ensemble, respectively~\cite{Ve94}. For SU(2)
and SU(3) gauge groups numerous results exist confirming the expected
relations~\cite{We99}.

We have investigated 4d U(1) gauge theory described by the action
$S \lbrace U_l \rbrace = \sum_p (1 - \cos \theta_p )$
with $U_l = U_{x,\mu} = \exp (i\theta_{x,\mu}) $ and
$
  \theta_p =
 \theta_{x,\mu} +
 \theta_{x+\hat{\mu},\nu} -
 \theta_{x+\hat{\nu},\mu} -
 \theta_{x,\nu}\ \ (\nu \ne \mu)\ . $
At $\beta_c \approx 1.01$ U(1) gauge theory undergoes a phase
transition between a confinement phase with mass gap and monopole
excitations for $\beta < \beta_c$ and the Coulomb phase which exhibits
a massless photon~\cite{BePa84} for $\beta > \beta_c$. The question of
the order of this phase transition, and hence the issue of a continuum
limit $\beta\to\beta_c-0$ of the massive phase, has remained a subject
of endless debate, see for instance~\cite{JeLaNe96} and references
therein. For $\beta > \beta_c$ a critical line of continuum theories
ought to exist as the photon is massless for all these $\beta$-values.

\begin{figure*}[t]
  \centerline{\psfig{figure=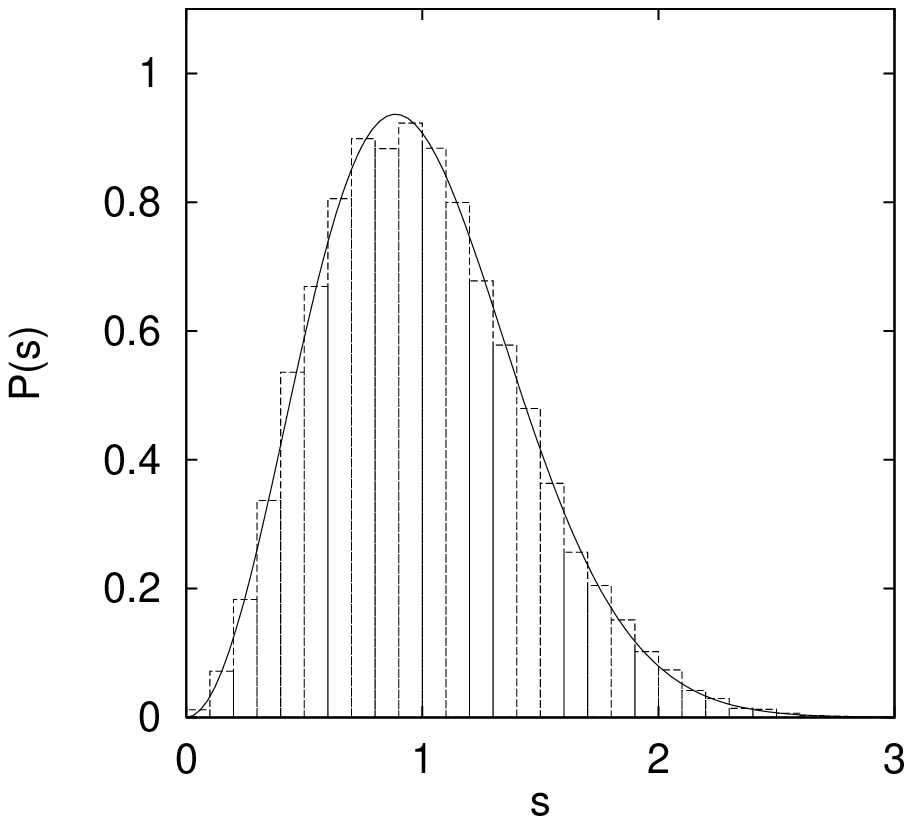,width=5cm}\hspace*{3mm}
    \psfig{figure=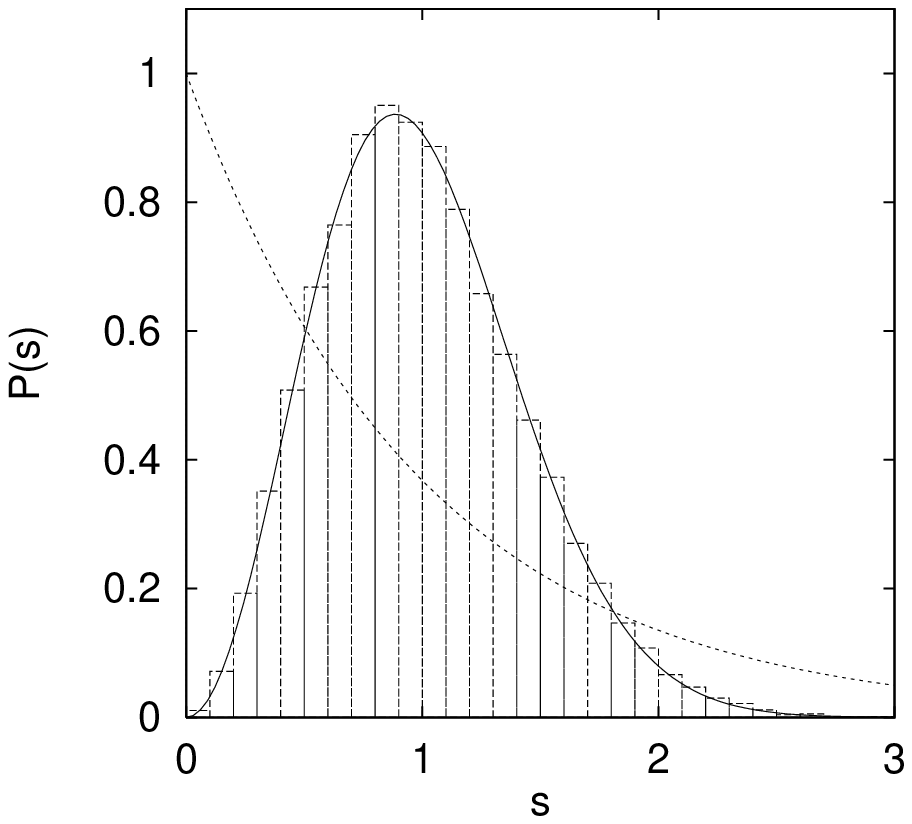,width=5cm}\hspace*{3mm}
    \psfig{figure=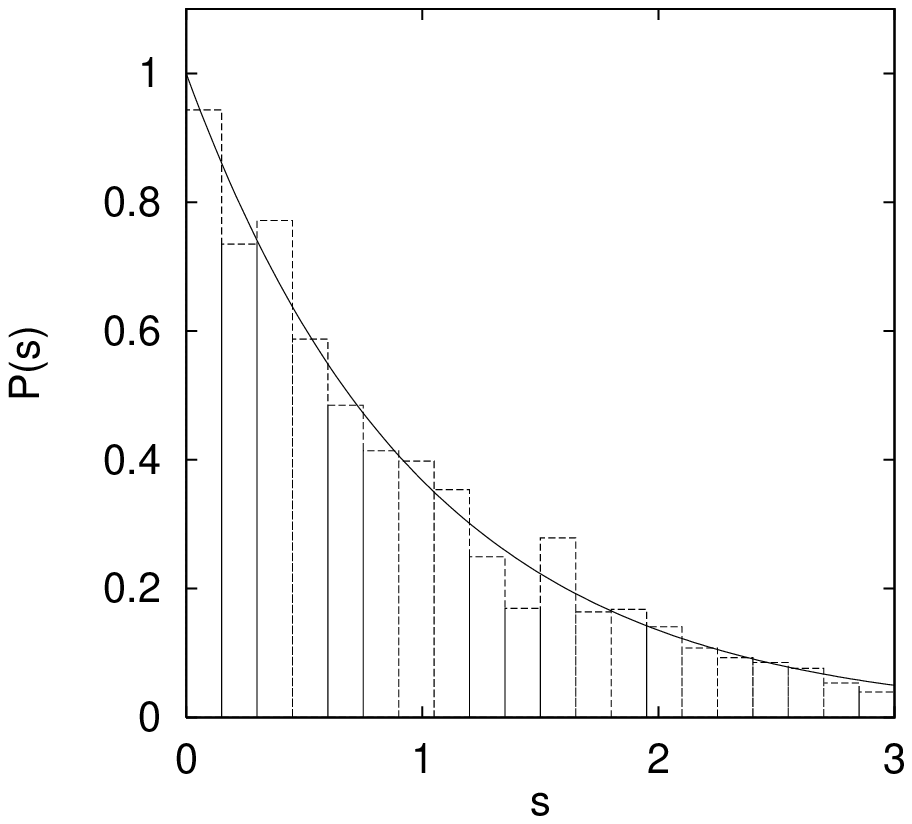,width=5cm}}
  \vspace*{-0.8cm}
  \caption{Nearest neighbor spacing distribution $P(s)$ on  an
    $8^3\times 6$ lattice in the confined phase (left plot) and in the
    Coulomb phase (central plot), and for the free Dirac operator on a
    $53\times 47\times 43\times 41$ lattice (right plot). The
    theoretical curves are the chUE result,
    $P(s)=32(s/\pi)^2\exp(-4s^2/\pi)$, and the Poisson distribution,
    $P(s)=\exp(-s)$.}  \vspace*{-0.3cm}
  \label{f02}
\end{figure*}

We are interested in the relationship between U(1) gauge theory and
RMT across this phase transition~\cite{BeMaPu98}. The
Bohigas-Giannoni-Schmit~\cite{BoGiSc84} conjecture states that quantum
systems whose classical counterparts are chaotic have spectral
fluctuation properties (measured, e.g., by the nearest-neighbor
spacing distribution $P(s)$ of unfolded eigenvalues) given by RMT,
whereas systems whose classical counterparts are integrable obey a
Poisson distribution, $P(s)=\exp (-s)$. As the Dirac operator is a
quantum-mechanical object without classical counterpart, the meaning
of the Bohigas-Giannoni-Schmit conjecture for lattice gauge theory is
somewhat unclear. Nevertheless, as for SU(2) and SU(3) gauge groups,
we expect the confined phase to be described by RMT, whereas free
fermions are known to yield the Poisson distribution. The question
arose whether the Coulomb phase will be described by RMT or by the
Poisson distribution, with apparently no clear theoretical arguments
in favor of either scenario.

In Ref.~\cite{BeMaPu98} some of the authors have resolved this
question by numerical analysis. We generated twenty (or more) gauge
configurations at each of the following parameter values: $8^3\times
4$ lattice at $\beta = 0$, $0.9$, $0.95$, $1$, $1.05$, $1.1$, $1.5$
and $8^3\times 6$ lattice at $\beta = 0.9$, $1.1$, $1.5$.  The
nearest-neighbor spacing distributions for the $8^3\times 6$ lattice
at $\beta=0.9$ (confined phase) and at $\beta=1.1$ (Coulomb phase),
averaged over 20 independent configuration, are depicted in
Fig.~\ref{f02}. Both are well described by the chUE. In contrast, the
right plot in Fig.~\ref{f02} shows that free fermions are described by
the Poisson distribution. The large prime numbers for the lattice size
are needed to remove the degeneracies of the spectrum.

We have continued the above investigation with a study of the
distribution of small eigenvalues in the confined phase. The
Banks-Casher formula~\cite{BaCa80} relates the eigenvalue density
$\rho(\lambda)$ at $\lambda=0$ to the chiral condensate,
\begin{equation} \label{Banks-Casher}
 \Sigma = |\langle \bar{\psi} \psi \rangle| =
 \lim_{m\to 0}\lim_{V\to\infty}
 \pi\rho (0)/V\:.
\end{equation}
The microscopic spectral density
\begin{equation} \label{rho_s}
 \rho_s (z) = \lim_{V\to\infty} {1\over V \Sigma}\,
 \rho \left( {z\over V\Sigma } \right)
\end{equation}
should be given by the result for the chUE of RMT~\cite{ShVe92}.  This
function also generates the Leutwyler-Smilga sum rules~\cite{LeSm92}.

\begin{figure}[tb]
  \begin{center}
  \begin{tabular}{rc}
   \raisebox{41mm}{$P(\lambda_{\min})$}
   & \hspace*{-3mm}\psfig{figure=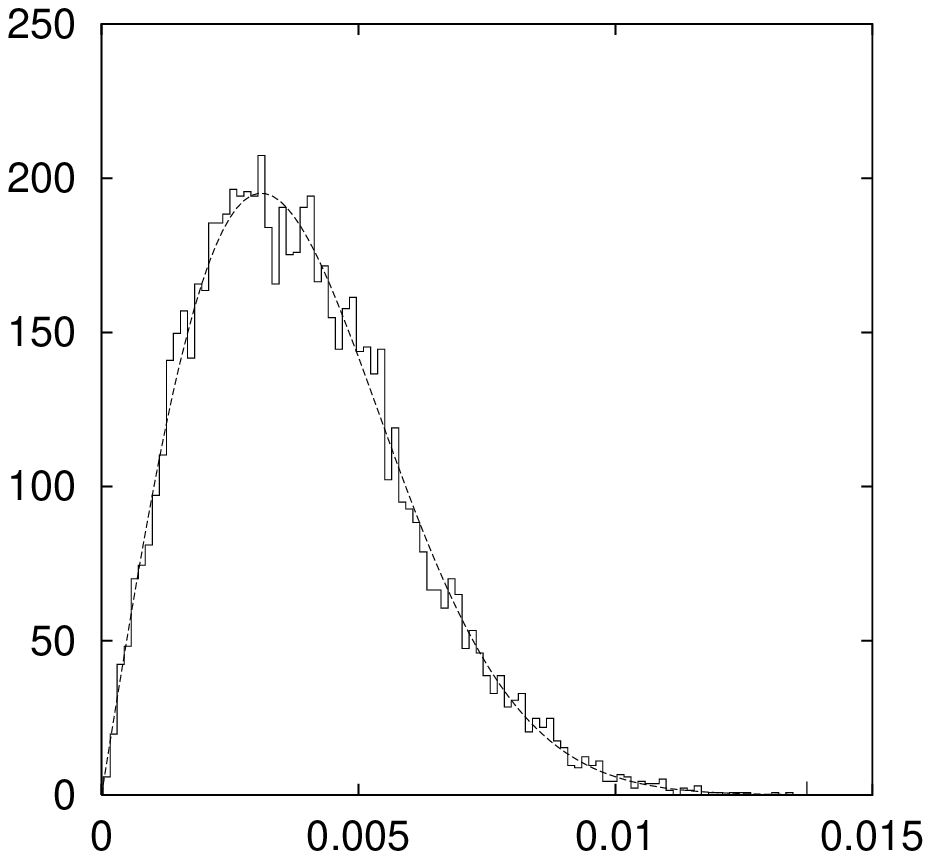,width=5.15cm}\\
   & \hspace*{32mm}$\lambda_{\min}$\\[3mm]
   \raisebox{41mm}{$\rho_s(z)$}
   & \hspace*{-4mm}\psfig{figure=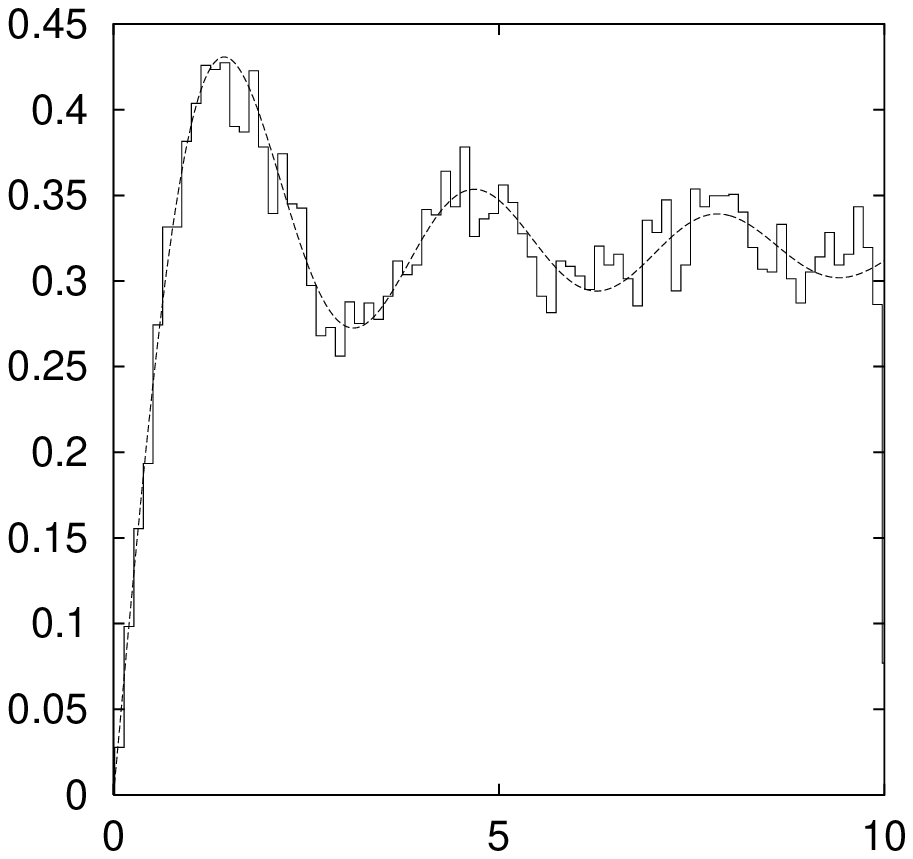,width=5cm}\\[-2mm]
   & \hspace*{25mm}$z$
  \end{tabular}
  \end{center}
  \vspace*{-12mm}
  \caption{Distribution $P(\lambda_{\min})$ (upper plot) and 
    microscopic spectral density $\rho_s (z)$ (lower plot) from our
    $6^4$ lattice data in comparison with the predictions of the chUE
    of RMT (dashed lines), see
    Eqs.~(\ref{plambdamin})~and~(\ref{rhosz}).} 
  \label{f06}
  \vspace*{-8mm}
\end{figure}

To study the smallest eigenvalues, spectral averaging is not possible,
and one has to produce large numbers of configurations. Our present
results are for $\beta=0.9$ in the confined phase with 10000
configurations on a $4^4$, 10000 configuration on a $6^4$, and 1106
configurations on an $8^3 \times 6$ lattice.  The upper plot in
Fig.~\ref{f06} exhibits the distribution $P(\lambda_{\min})$ of the
smallest eigenvalue $\lambda_{\min}$ in comparison with the prediction
of the (quenched) chUE of RMT for topological charge $\nu=0$,
\begin{equation}
\label{plambdamin}
P(\lambda_{\min}) = {(V\Sigma)^2 \lambda_{\min} \over 2}\,\exp\left( - 
{(V\Sigma\lambda_{\min})^2 \over 4} \right).
\end{equation}
The agreement is excellent for all lattices.  For the chiral
condensate we obtain $\Sigma \approx 0.35$ by extrapolating the
histogram for $\rho(\lambda)$ to $\lambda=0$ and using
Eq.~(\ref{Banks-Casher}). (With staggered fermions on a finite lattice
one always has $\rho(0)=0$, but within reasonable limits the
extrapolated value is independent of the choice of the bin size.)
Since the average value of $\lambda_{\min}$ goes like $V^{-1}$,
$\langle\lambda_{\min}\rangle$ decreases with increasing lattice size.
In the lower plot of
Fig.~\ref{f06} the same comparison with RMT is done for the microscopic
spectral density~(\ref{rho_s}) up to $z=10$, and the agreement is
again quite satisfactory. Here, the analytical RMT result for the
(quenched) chUE and $\nu=0$ is given by
\begin{equation}\label{rhosz}
\rho_s(z) = {z \over 2} \, [ J_0^2(z) + J_1^2(z) ] \,.
\end{equation}

\begin{figure}[tb]
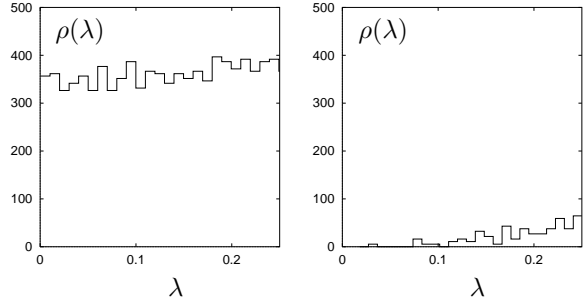

  \hspace*{6mm}\vspace*{-6mm}$\rho(\lambda)$\hspace*{33.7mm}$\rho(\lambda)$
  \centerline{\psfig{figure=f12.eps,width=3.5cm}\hspace*{4mm}
    \psfig{figure=f11.eps,width=3.5cm}}
  \vspace*{-2mm}
  \hspace*{21mm}$\lambda$\hspace*{38mm}$\lambda$
  \vspace*{-8mm}
  \caption{Density $\rho(\lambda)$ of small eigenvalues for the $8^3\times 6$ lattice
    at $\beta = 0.9$ (left plot) and at $\beta = 1.1$ (right plot). A
    non-zero chiral condensate is supported in the confinement phase.}
  \label{f12}
  \vspace*{-0.5cm}
\end{figure}

The quasi-zero modes which are responsible for the chiral condensate
$\Sigma \approx 0.35$ build up when we cross from the Coulomb into the
confined phase. For our $8^3\times 6$ lattice, Fig.~\ref{f12} compares
on identical scales densities of the small eigenvalues at $\beta =
0.9 $ (left plot) and at $\beta = 1.1$ (right plot), averaged over
20 configurations. The quasi-zero modes in the left plot
are responsible for the non-zero chiral
condensate $\Sigma>0$ via Eq.~(\ref{Banks-Casher}), whereas no such
quasi-zero modes are found in the Coulomb phase. This is as expected.
However, it may be worthwhile to understand the physical origin of the
U(1) quasi-zero modes in more detail. For 4d SU(2) and SU(3) gauge
theories a general interpretation is to link them, and hence the
chiral condensate, to the existence of instantons. As there are no
instantons in 4d U(1) gauge theory, one needs another explanation, and
it may be interesting to study similarities and differences to the 4d
SU(2) and SU(3) situations. An analogous case exists in 3d QCD
\cite{VeZa93b}.

In conclusion, the nearest-neighbor spacing distribution of 4d U(1)
quenched lattice gauge theory is described by the chUE in both the
confinement and the Coulomb phase.  In the confinement phase we also
find that the $P(\lambda_{\min})$ distribution and the microscopic
spectral density (\ref{rho_s}) are described by the chUE.  A better
physical understanding of the origin of the quasi-zero modes, which are
responsible for the non-zero chiral condensate, is desirable.


\begin{thebibliography}{19}
  
\bibitem{ShVe92} E.V. Shuryak and J.J.M. Verbaarschot, Nucl. Phys.
  A560 (1992) 306; J.J.M. Verbaarschot and I. Zahed, Phys. Rev. Lett.
  70 (1993) 3852; J.C. Osborn, D. Toublan, and J.J.M. Verbaarschot,
  Nucl. Phys. B540 (1999) 317; P.H. Damgaard, J.C. Osborn, D. Toublan,
  and J.J.M. Verbaarschot, Nucl. Phys. B547 (1999) 305.

\bibitem{Ve94} J.J.M. Verbaarschot, Phys. Rev. Lett. 72 (1994) 2531.

\bibitem{We99} see, e.g., T. Wettig, hep-lat/9905020.

\bibitem{BePa84} B.A. Berg and C. Panagiotakopoulos, Phys. Rev. Lett.
52 (1984) 94.

\bibitem{JeLaNe96} J. Jers\'ak, C.B. Lang, and T. Neuhaus,
Phys. Rev. Lett. 77 (1996) 1933.

\bibitem{BeMaPu98} B.A. Berg, H. Markum, and R. Pullirsch, 
Phys. Rev. D59 (1999) 097504.

\bibitem{BoGiSc84} O. Bohigas, M.-J. Giannoni, and C. Schmit, Phys.
Rev. Lett. 52 (1984) 1.

\bibitem{BaCa80} T. Banks and A. Casher, Nucl. Phys. B169 (1980) 103.

\bibitem{LeSm92} H. Leutwyler and A.V. Smilga, Phys. Rev. D46 (1992)
5607.

\bibitem{VeZa93b} J.J.M. Verbaarschot and I. Zahed, Phys. Rev. Lett.
73 (1994) 2288.

\end{thebibliography}
\end{document}